# Advanced High-Performance Large Diameter Cs$_2$HfCl$_6$ (CHC) and Mixed Halides Scintillator

*Rastgo Hawrami[1*], Elsa Ariesanti[1], Vlad Buliga,[1] Liviu Matei,[1] Shariar Motakef[2], Arnold Burger[1]*

[1]Fisk University, Nashville, TN 37208 USA

[2]CapeSym Inc., Natick, MA 01760 USA

ABSTRACT

      This paper reports on successful growth and performance evaluation of two large diameter Cs$_2$HfCl$_6$ (CHC) and Cs$_2$HfCl$_4$Br$_2$ (CHCB), both recently developed scintillator crystals. The discovery of Cs$_2$HfCl$_6$ (CHC) as a scintillator has lately generated much interest in this material and its family, which belongs to the K$_2$PtCl$_6$ cubic crystal structure. CHC is an intrinsic scintillator that is non-hygroscopic, has no self-radioactivity, provides excellent energy resolution, and has

[*] Corresponding Author: email: drh1980@gmail.com, phone: 1-615-916-6666.

excellent non-proportionality. CHC has a moderate density of 3.9 g/cm$^3$ and an effective atomic number of 58. Reported in this paper are transparent crack-free single crystal CHC and CHCB boules of one inch in diameter, both grown using the vertical Bridgman method. Samples retrieved from the boules, sized ⌀23mm × 30mm and ⌀23mm × 26mm, respectively, are characterized for their optical and scintillation properties. Energy resolutions of 3.5% and 3.7% (FWHM) at 662 keV, respectively, are reported. Results comparable to previously reported results for smaller crystals have been obtained. Studies on light yield, decay time, non-proportionality, as well as detector characterization are also reported.

INTRODUCTION

Widespread use of scintillators as gamma-ray detectors is largely generated by their tunable properties and extensive availability. Currently there are needs for scintillators with targeted properties such as high light output, high stopping power ($Z_{eff}$), fast decay time, good linearity and low cost. The most commonly used scintillators like NaI:Tl and CsI:Tl are widely used in gamma cameras for medical imaging. Recently, there has been a growing interest in employing a large volume of scintillator material such as NaI:Tl or CsI:Tl in detection and identification of radioisotopes for homeland security. Different radioisotopes can have close emission energies, which need to be distinguishable for accurate detection. Therefore, better performance in distinguishing between the different isotopes will require the scintillation material to have excellent energy resolution, excellent proportionality, and high light yield. Newly discovered cerium-doped lanthanum halide scintillators, such as LaCl$_3$:Ce, and LaBr$_3$:Ce, are well known to have excellent energy resolutions (about 3% (FWHM) at 662 keV) and have a more linear response to energy than NaI:Tl. However, but they have intrinsic radioactivity and most of past measurements were performed using small research crystals. The market for scintillation detectors annually consumes tens of tons of alkali halide scintillation crystals. This being so, new application fields arise along with traditional ones. They dictate a necessity for R&D of new technologies for more perfect crystal growth. Thus, the growth of alkali halide crystals is a core matter of scientific and technological activity in production and improvement of alkali halide scintillators[1].

Since the re-discovery of CHC as a new non-hygroscopic simple cubic and intrinsic scintillator, a few papers have reported their studies on the growth difficulties. CHC was re-discovered and reported as an example of a little-known class of non-hygroscopic compounds having the generic cubic crystal structure of $K_2PtCl_6$[2]. CHC scintillation is centered at 400 nm, with a principal decay time of 4.4 µs and a light yield of up to 54,000 photons/MeV (when compared to BGO), and energy resolution of 3.3% at 662 keV using a 0.65 cm$^3$ cubic sample[3]. A further study on crystal growth and behavior CHC and its variant, $Cs_2HfCl_4Br_2$ (CHCB) was also reported[4]. CHC and CHCB crystals were prepared by melt compounding of sublimed $HfCl_4$ with CsCl and CsBr to produce material for Bridgman growth. A clear CHC sample had a light yield and energy resolution of 30,000 ph/MeV (when compared to NaI:Tl) and 3.3%, respectively, with decay components of 0.39 and 3.9 µs. A sample of CHCB with a secondary phase present in the core had a light yield and energy resolution of 18,600 ph/MeV and 4.4%, and with decay components of 0.38 and 2.0 µs for CHCB. Both crystals showed minimal moisture sensitivity[4]. In both previous papers small samples were presented and tested due to difficulties in crystal growth[4]. Shown in their resulting ⌀1-cm and ⌀1-inch CHC crystals is evidence of CsCl as a secondary phase, which was a result of a non-stoichiometric (CsCl-rich) melt composition caused by the high vapor pressure of $HfCl_4$ during compounding. This secondary phase was verified using micro X-ray fluorescence spectrometry[4]. The study also reported an evidence of a secondary phase in CHCB. In this paper we are presenting growth and characterization results of high performing transparent crack-free large diameter CHC and CHCB single crystals with energy resolutions of 3.5% and 3.7%, light yields of 23,000 ph/MeV and 20,000 ph/MeV, and primary decay times of 4.3 µs and 1.8 µs, respectively, which indicates that CHCB is close to 2.5 times faster than CHC.

EXPERIMENTAL METHODS

For a growth run stoichiometric amounts of starting materials CsCl (99.999%) and $HfCl_4$ (99.9%) for CHC, along with CsCl, CsBr (99.995%) and $HfBr_4$ (99.9%) for CHCB, were utilized. Prior to growth, $HfCl_4$ starting material underwent a one-time purification by sublimation. For this purification process 200 grams of $HfCl_4$ powder was loaded into a pre-cleaned 1-inch inner diameter quartz ampoule, which was subsequently sealed under vacuum. The ampoule was inserted into a one-zone horizontal furnace, with the furnace temperature set to 220ºC. The sublimation process was conducted for 72 hours, after which the furnace was cooled down to room temperature. Figure 1 shows the sublimation ampoule post-processing, where about 40% of $HfCl_4$ was left at the starting end (left-side) of the ampoule.

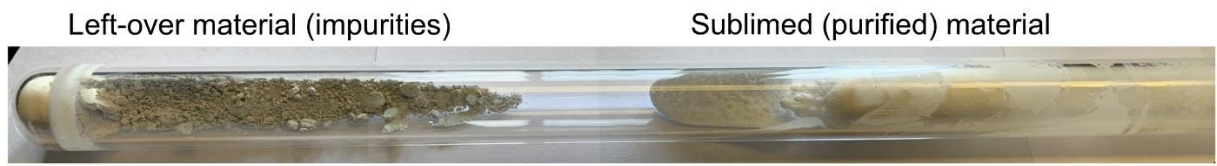

**Figure 1**. Result of $HfCl_4$ sublimation in the purification ampoule. The sublimed or purified $HfCl_4$ is shown in the right side of the ampoule, while the leftover material is shown on the left.

For a growth run stoichiometric amounts of starting materials CsCl and purified $HfCl_4$ for CHC, along with CsCl, CsBr and $HfBr_4$ for CHCB, were loaded into different sizes growth ampoules such as ∅16mm and a ∅ one-inch ampoules. The ampoules were then attached to a vacuum system until a high vacuum ≤1 × 10$^{-5}$ Torr was reached. The ampoules were then sealed and subsequently placed into a two-zone furnace for growth by the vertical Bridgman method (Figure 2). The top furnace zone temperature was set a few degrees above melting point of CHC and CHCB at 800°C and the bottom furnace zone a few degrees below the melting point. The growth rate of 3-4 mm/hour was used. When the growth cycle was completed both top and bottom zones were cooled at a rate of 100 to 150°C/day down to room temperature.

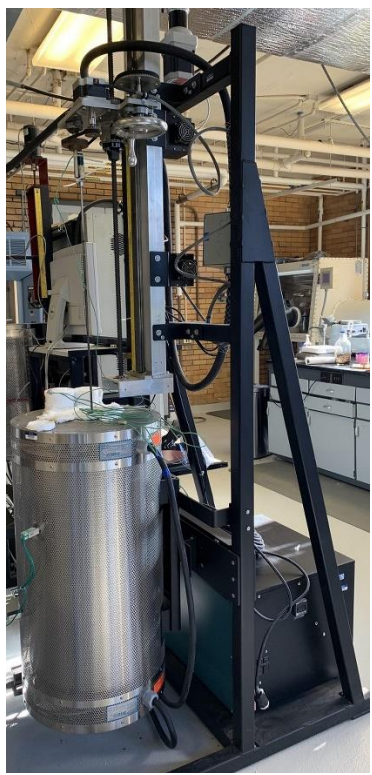

**Figure 2.** A two-zone furnace for growth by the vertical Bridgman method.

Figures 3(a) and 3(b) show photographs of as-retrieved ∅16mm and ∅1-inch CHC boules, respectively, which were successfully grown single, crack-free, inclusion-free crystals. Results from ∅16 mm boule were reported[5] and are being included in another publication[6]. From the and ⌀1-inch CHC boule a 31 mm long sample was cut using a diamond wire saw for evaluation. Figure 4(a) shows the transparent and crack-free CHC sample, now with dimensions of ⌀23 mm × 30 mm after lapping and polishing. Two samples from a ⌀1-inch CHCB boule are shown in Figure 4(b). A roughly lapped ⌀23 mm × 26 mm cylinder (left), prior to being finely polished, and a thinner finely polished (up to 4000 grit) plate (right) show that the grown ⌀1-inch CHCB crystal boule was clear and crack-free.

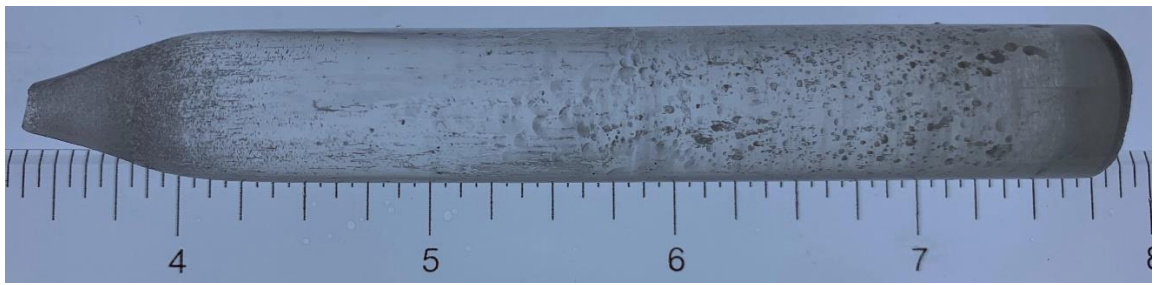

**Figure 3. (a)** As-retrieved grown ⌀ 16 mm CHC boule (inch ruler is shown).

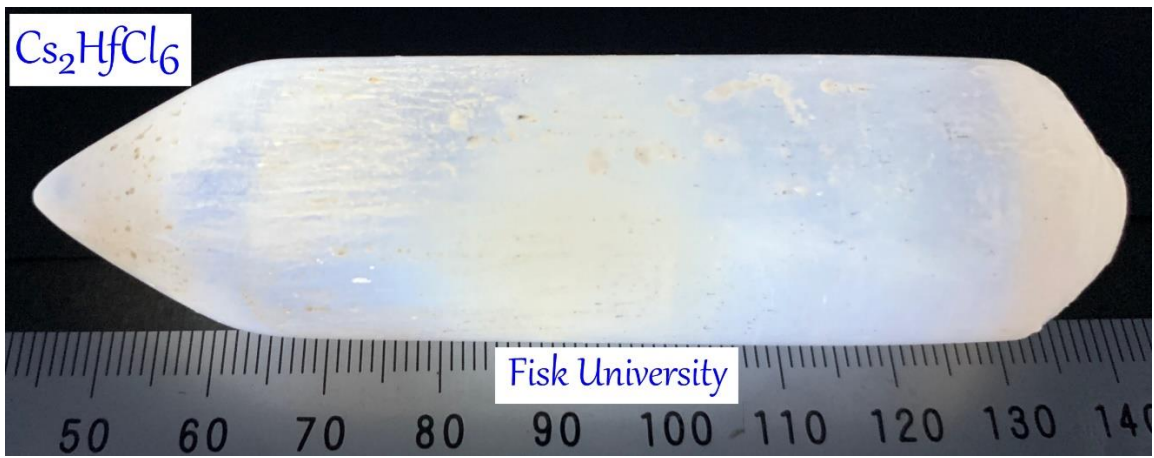

**Figure 3. (b)** As-retrieved grown ⌀ 1-inch CHC boule (cm ruler is shown).

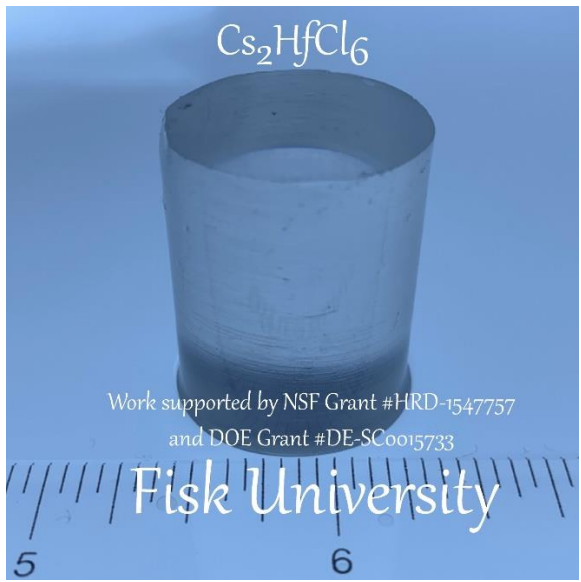

**Figure 4**. **(a)** CHC sample cut from a ⌀1-inch boule, with dimensions of ⌀23 mm × 30 mm after lapping and polishing.

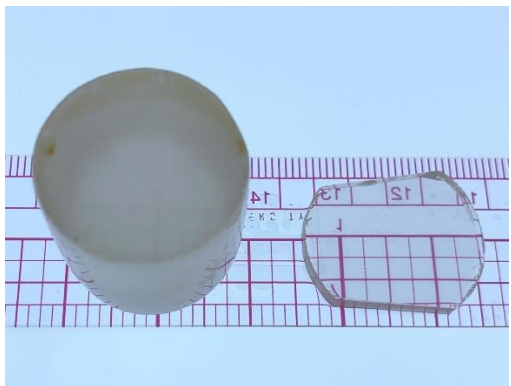

**Figure 4**. **(b)** Two samples from a ⌀1-inch CHCB boule, a roughly lapped ⌀23 mm × 26 mm cylinder (left) and prior to being finely polished, as well as a thinner, finely polished (up to 4000 grit) plate (right).

Intrinsic detection efficiency as a function of γ-ray energy (in MeV) for a ⌀1″× 1″ CHC compared to that of ⌀1″× 1″ $SrI_2$, NaI, and $LaBr_3$ (Figure 4) shows one of the motivations for pursuing growth of large diameter crystals of CHC (and CHCB). As seen in Figure 4 up to 100 keV, all four scintillators have the same intrinsic detection efficiency. For any γ-ray energy above 100 keV, NaI has the lowest intrinsic detection efficiency compared to the other three scintillators. While for this energy range CHC has also a slightly lower efficiency compared to LaBr$_3$ and SrI$_2$, CHC is both not doped and not hygroscopic, advantages to the other three scintillators.

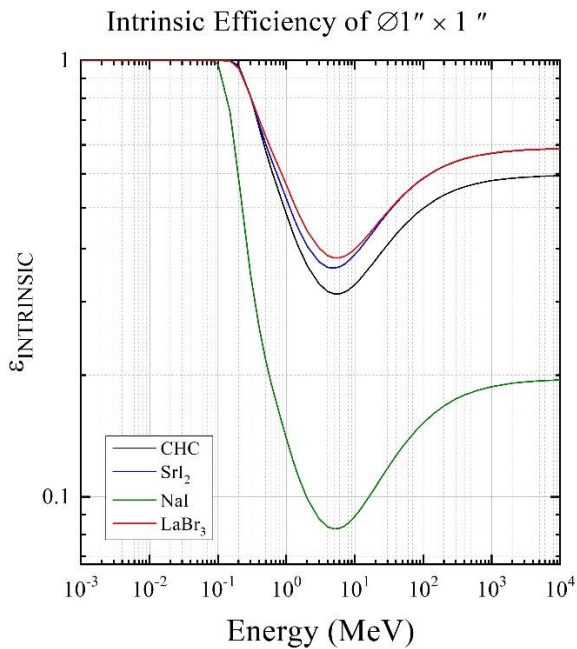

**Figure 5.** Intrinsic efficiency calculation for CHC, compared to SrI$_2$, NaI, and LaBr$_3$, each with dimensions ∅1″ × 1″. Mass attenuation coefficient data were generated using the online NIST XCOM software[7].

The response of CHC and CHCB samples to gamma-rays was measured by placing a crystal in oil within a quartz cup lined with Teflon tape as a diffuse reflector. The oil cup was coupled to Hamamatsu R6231-100 superbialkali photo-multiplier tube (PMT) using Saint Gobain BC630 optical grease. Through a voltage divider the PMT was powered by Hamamatsu

C9520-02 power supply at -700V. Signals from the PMT were processed through Canberra 2005 spectroscopic amplifier, Canberra 2022 amplifier (with 180X gain and 12 µs shaping time), and finally analyzed by Amptek MCA8000D multichannel analyzer. Differential height spectra from check sources with x-/γ-ray energies between 14 keV and 1332 keV were collected with CHC and CHCB samples to obtain energy resolution and light yield information at specific photon energies. Relative light yield data were obtained by normalizing the light yield at the specific energy to that of $^{137}$Cs. The non-proportionality curve was obtained by plotting the relative light yield as a function of photon energy. Decay time data was obtained by analyzing the signals collected at the PMT output with CAEN D5720C digitizer. The averaged signals were analyzed offline and fitted with an exponential function with two decay times constants.

RESULTS AND ANALYSIS

Figure 6(a) shows a $^{137}$Cs spectrum collected with a ⌀23 mm × 30 mm CHC sample shown in Figure 3(a), showing an energy resolution of 3.5% (FWHM) at 662 keV that is comparable to ≤3.0% (FWHM) previously measured for a smaller size ⌀15mm × 15mm CHC crystal[6,7]. The light yield was measured to be 23,000 ph/MeV, by comparing the 662 keV full energy peak channel number for CHC with that of a ⌀1″×1″ NaI:Tl (Figure 6(b)) while taking into account the quantum efficiency of R6321-100 PMT at 410 nm (emission peak of CHC[3]) and at 415 nm (emission peak of NaI:Tl[8]). The light yield value, slightly lower than previously reported for smaller and thinner CHC crystals, may be attributed to non-doped (intrinsic) scintillation and/or less than optimal light collection. Thinner (i.e. height < diameter) CHC crystals have been observed to produce higher light yield than thicker (i.e. height diameter) crystals. Factors that influence light collection such as index of refraction and light yield collection efficiency calculation for CHC will be explored in a future publication.

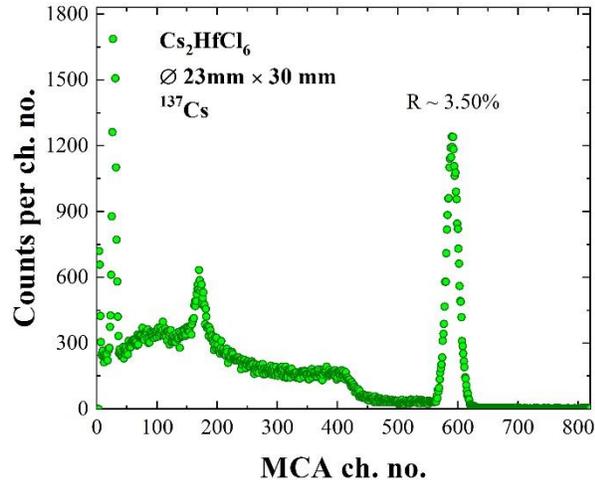

**Figure 6**. **(a)** $^{137}$Cs spectrum collected with CHC sample shown in Figure 4(a). Energy resolution of 3.5% (FWHM) at 662 keV was measured.

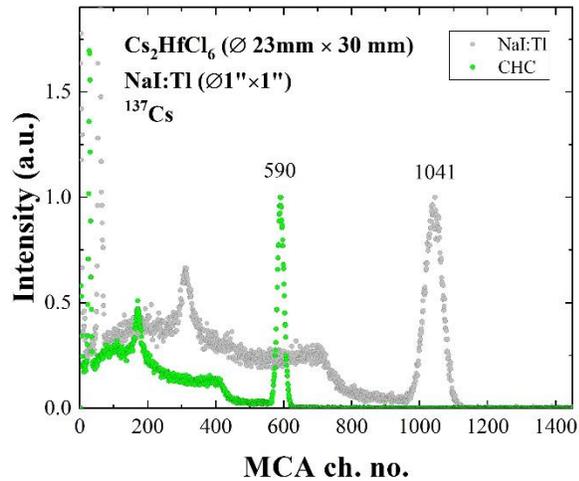

**Figure 6**. **(b)** Comparison of (a) with $^{137}$Cs spectrum collected with a ⌀1″×1″ NaI:Tl, resulting in a light yield of 23,000 ph/MeV.

$^{137}$Cs spectrum collected the CHCB sample in Figure 4(b) is shown in Figure 7(a), showing an energy resolution of 3.7% (FWHM) at 662 keV. The light yield, calculated with the same procedure as with CHC and with taking into account the quantum efficiency of the PMT at 423 nm (the emission peak for CHCB[9]), is 20,000 ph/MeV (Figure 7(b)). Better crystal quality contributes to a better energy resolution (compared to previously reported values[4,9]). Like the case of large diameter CHC crystals, large diameter CHCB crystals may benefit from optimization of light collection efficiency.

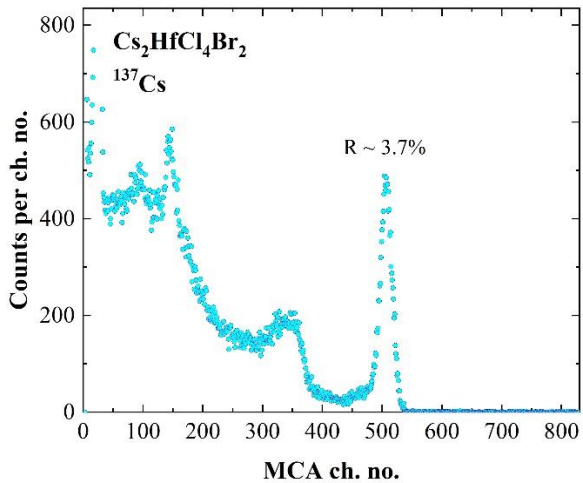

**Figure 7**. **(a)** $^{137}$Cs spectrum collected with CHCB sample shown in Figure 4(a). Energy resolution of 3.7% (FWHM) at 662 keV was measured.

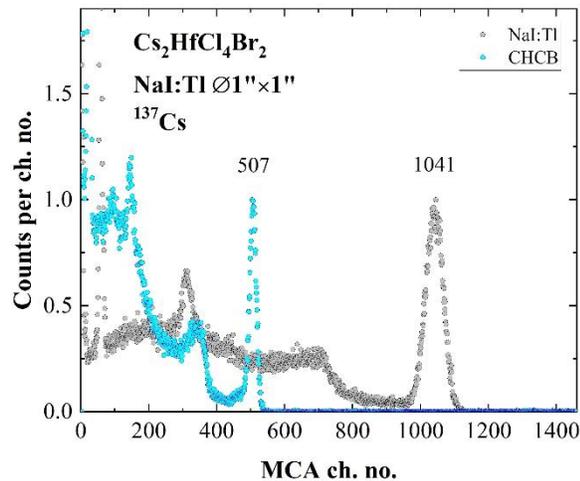

**Figure 7**. **(b)** Comparison of (a) with $^{137}$Cs spectrum collected with a ∅1″×1″ NaI:Tl, resulting in a light yield of 20,000 ph/MeV.

Spectra from check sources with x-/γ-ray energies between 14 keV and 1332 keV were collected to obtain full energy peak resolution and light yield at specific photon energies. Relative light yield data were obtained by normalizing the light yield at the specific energy to that of $^{137}$Cs. The γ-ray non-proportionality curves for CHC in Figure 7(a) and for CHCB in Figure 7(b) were obtained by plotting the relative light yield as a function of photon energy. Also plotted in Figures 7(a) and 7(b) are the non-proportionality curves for a ∅1″×1″ NaI:Tl and a 1cm$^3$ BGO for comparison. Each data set were fitted with a fourth order polynomial curves for a better data trend observation. For energy above 30 keV both CHC and CHCB have a fairly linear response to γ-rays (i.e., relative light yield deviates less than ±5% from unity). Such behavior was also observed for smaller CHC and CHCB crystals and was also observed from CHC's electron non-proportionality measurement by SLYNCI[11].

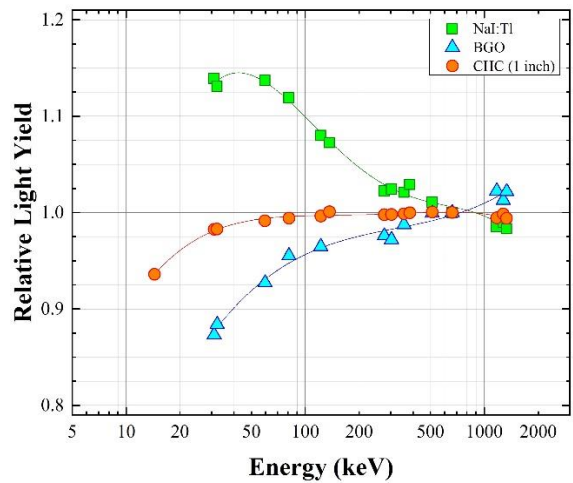

**Figure 8**. **(a)** Relative light yield as a function of γ-ray energy for CHC is compared to that NaI:l and BGO, showing that for γ-rays above 30 keV CHC has a fairly linear response.

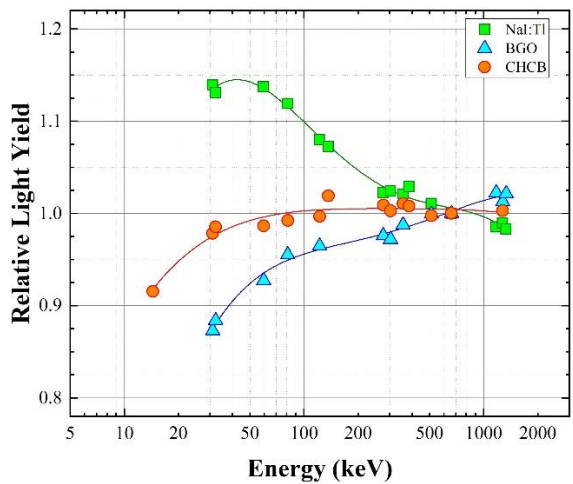

**Figure 8**. **(b)** Relative light yield as a function of γ-ray energy for CHCB is compared to that NaI:l and BGO, also showing a fairly linear response above 30 keV.

Decay time data was obtained by analyzing the signals collected at the PMT output with CAEN D5720C digitizer. The averaged signals were analyzed offline and fitted with an exponential function with two decay times constants. The average of the waveforms for CHC and CHCB are plotted and shown in Figures 9(a) and 9(b), respectively. Fitting of the averaged waveforms with an exponential function with two decay time constants determines the decay times of the CHC sample in Figure 3(a) to be 254 ns (7%) and 3.8 μs (93%), which are comparable to published data from smaller CHC crystals[3,4,6,7]. With similar fitting procedure, the decay times for CHCB were found to be 330 ns (10%) and 1.8 μs (90%), values which are also comparable to published data from smaller CHCB crystals[4,9]. Substituting some Cl$^-$ ions with Br$^-$ ions appears to halve the primary decay time from 3.8 μs to 2 μs. Because CHC and CHCB are both intrinsic scintillators, the improvement to scintillation timing property may be caused by the change in crystal structure or cell parameters, and possibly the change in energy band diagram. A theoretical study on the effects of Br- ion substitutions in CHC to the crystal structure, like the one already conducted for CHC[11], is currently being carried out and will be reported in a future publication.

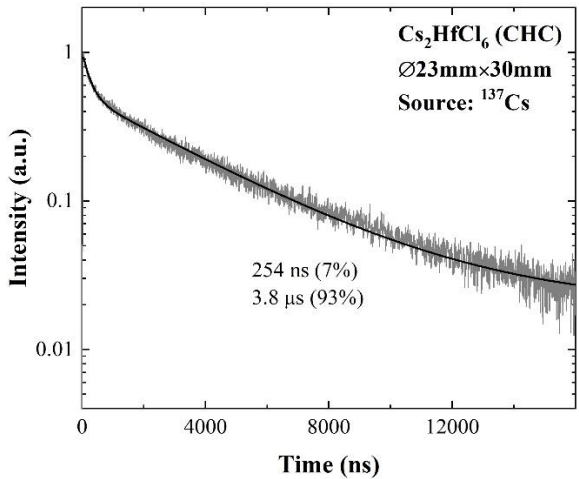

**Figure 9. (a)** Decay times of CHC, measured to be 254 ns (7%) and 3.8 μs (93%).

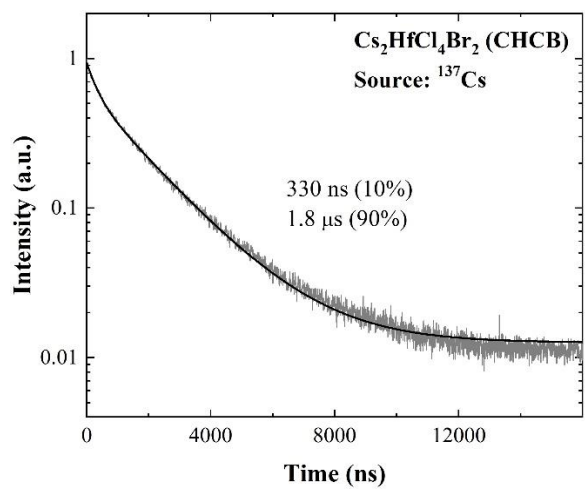

**Figure 9. (b)** Decay times of CHC, measured to be 330 ns (10%) and 1.8 μs (90%).

CONCLUSIONS

Cs$_2$HfCl$_6$ is a non-hygroscopic, simple cubic intrinsic high performance scintillator crystal. We are reporting a development of a high performing transparent crack-free CHC single crystal growth that is achievable using highly purified in lab processed starting materials, specifically HfCl$_4$. Half inch and one-inch diameter CHC and CHCB crystals have excellent energy resolutions about 3.5% (FWHM) at 662 keV and proportional response to photon energy over a large range of photon energies. Larger diameter CHC and CHCB samples perform as well as smaller CHC crystals indicating high crystal quality. The primary scintillation decay constant for CHCB is close to 2.5 times faster than CHC.

AUTHOR INFORMATION

**Corresponding Author**

*e-mail: drh1980@gmail.com, phone: 1-615-916-6666.

**Author Contributions**

The manuscript was written through contributions of all authors. All authors have given approval to the final version of the manuscript.

**Funding Sources**

The research in this manuscript was partly funded by U.S. Department of Energy under Grant #DE-SC0015733, U.S. National Science Foundation under Grant #HRD-1547757, and U.S. National Aviation and Space Agency under Grant NNX16AK42G.

ACKNOWLEDGMENT

The research in this manuscript was partly funded by U.S. Department of Energy under Grant #DE-SC0015733, U.S. National Science Foundation under Grant #HRD-1547757, and U.S. National Aviation and Space Agency under Grant NNX16AK42G.

ABBREVIATIONS

CHC, $Cs_2HfCl_6$; CHCB, $Cs_2HfCl_4Br_2$; NIM, Nuclear Instrument Module; PMT, photomultiplier tube; SLYNCI, Scintillator Light Yield Non-proportionality Characterization Instrument